\documentclass[sigconf]{acmart}

\usepackage{soul}
\usepackage{siunitx}
\usepackage{bbding}
\usepackage{pifont}

\AtBeginDocument{%
  \providecommand\BibTeX{{%
    \normalfont B\kern-0.5em{\scshape i\kern-0.25em b}\kern-0.8em\TeX}}}

\copyrightyear{2022}
\acmYear{2022}
\setcopyright{acmlicensed}\acmConference[CVMP '22]{European Conference on Visual Media Production}{December 1--2, 2022}{London, United Kingdom}
\acmBooktitle{European Conference on Visual Media Production (CVMP '22), December 1--2, 2022, London, United Kingdom}
\acmPrice{15.00}
\acmDOI{10.1145/3565516.3565522}
\acmISBN{978-1-4503-9939-5/22/12}

\renewcommand\footnotetextcopyrightpermission[1]{} 
\begin{document}

\title[Tragic Talkers Dataset]{Tragic Talkers: A Shakespearean Sound- and Light-Field Dataset for  Audio-Visual Machine Learning Research}

\author{Davide Berghi}
\orcid{0000-0001-6279-6364}
\affiliation{%
  \institution{CVSSP, University of Surrey}
  \city{Guildford}
  \country{U.K.}
}
\authornote{This is the author's version of the work. It is posted here for your personal use. This paper is published under a Creative Commons Attribution (CC-BY) license. The definitive version was published in CVMP '22, https://doi.org/10.1145/3565516.3565522.}
\email{d.berghi@surrey.ac.uk}

\author{Marco Volino}
\orcid{0000-0002-1869-9257}
\affiliation{%
  \institution{CVSSP, University of Surrey}
  \city{Guildford}
  \country{U.K.}}
\email{m.volino@surrey.ac.uk}

\author{Philip J. B. Jackson}
\orcid{0000-0001-7933-5935}
\affiliation{%
  \institution{CVSSP, University of Surrey}
  \city{Guildford}
  \country{U.K.}}
\email{p.jackson@surrey.ac.uk}


\begin{abstract}

  3D audio-visual production aims to deliver immersive and interactive experiences to the consumer. Yet, faithfully reproducing real-world 3D scenes remains a challenging task. This is partly due to the lack of available datasets enabling audio-visual research in this direction.   
  In most of the existing multi-view datasets, the accompanying audio is neglected. Similarly, datasets for spatial audio research primarily offer unimodal content, and when visual data is included, the quality is far from meeting the standard production needs.     
  We present ``Tragic Talkers'', an audio-visual dataset consisting of excerpts from the ``Romeo and Juliet'' drama captured with microphone arrays and multiple co-located cameras for light-field video. Tragic Talkers provides ideal content for object-based media (OBM) production. It is designed to cover various conventional talking scenarios, such as monologues, two-people conversations, and interactions with considerable movement and occlusion, yielding 30 sequences captured from a total of 22 different points of view and two 16-element microphone arrays.
  Additionally, we provide voice activity labels, 2D face bounding boxes for each camera view, 2D pose detection keypoints, 3D tracking data of the mouth of the actors, and dialogue transcriptions.  
  We believe the community will benefit from this dataset as it can assist multidisciplinary research. Possible uses of the dataset are discussed.
\end{abstract}



\begin{teaserfigure}
  \includegraphics[width=\textwidth]{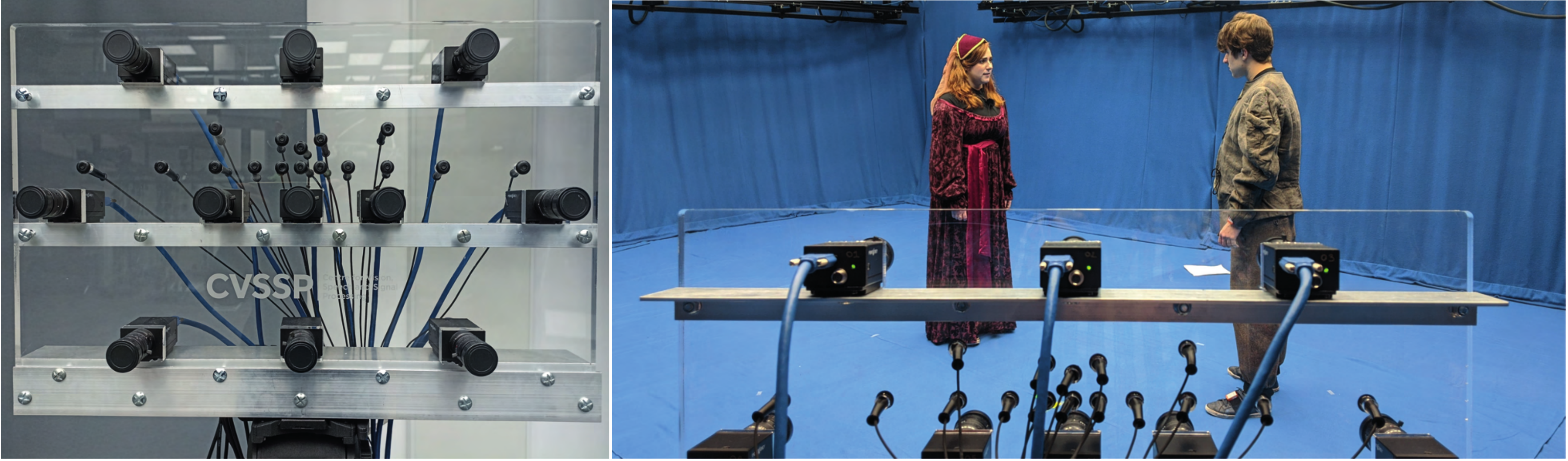}
  \caption{On the left, a photo of one of the two AVA Rigs used to capture the Tragic Talkers dataset. On the right, a photo taken from behind one of the rigs during the dataset capture session.}
  \Description{A frontal photo of the AVA Rig; and a photo taken from behind one AVA Rig while capturing the actors playing Romeo and Juliet in the blue studio. }
  \label{fig:teaser}
\end{teaserfigure}

\maketitle

\begin{table*}[bt]
  
  \caption{Summary of existing audio-visual datasets captured with a microphone array or binaural microphones for speech-related research. `Col.': co-located platform; `\# m': number of microphones in the microphone array; `$f_s$': audio sampling frequency in kHz; `\# c': number of cameras; `Res.': video resolution; `Cal.': camera calibration data; `BG': chroma-uniform background; `Act.': actors employed; `int.': intrinsics-only; `VA': voice activity annotations; `pos.': position; `transc.': transcripts}
  \label{tab:datasets}
  \begin{tabular}{cccccccccl}
    \toprule
    Dataset & Col. & \# m & $f_s$ & \# c & Res. & Cal. & BG & Act. & Labels\\
    \midrule
    AV16.3 & \ding{56} & 8($\times$2) & 16 & 3 & 360$\times$288 & \Checkmark & \ding{56} & \ding{56} & VA, 3D mouth and head pos.$^{a}$ \\
    \cite{Lathoud:2004:AV16} \\
    CAVA & \Checkmark & 2 & 44.1 & 2+1 & 1024$\times$768 & \Checkmark & \ding{56} & \ding{56} & Corner points \\
    \cite{Arnaud:2008:CAVA} \\
    AVDIAR & \Checkmark & 6 & 48 & 2 & 1920$\times$1200 & \Checkmark & \ding{56} & \ding{56} & VA, 2D bbox \\
    \cite{Gebru:2018:SpeakerDiarization} \\
    SSLR$^{b}$ & \Checkmark & 4 & 48 & 1 & 640$\times$480 & \Checkmark & \ding{56} & \ding{56} & VA, 3D tracking data \\
    \cite{He:2018:deepNetsHRI} \\
    CAV3D & \Checkmark & 8 & 96 & 1 & 1024$\times$768 & \Checkmark & \ding{56} & \ding{56} & VA, 3D mouth pos. \\
    \cite{Qian:2019:AVSensDevic} \\
    EasyCom & \Checkmark & 6 & 48 & 1 & 1920$\times$1080 & int. & \ding{56} & \Checkmark & VA, head pose, 2D face+head bbox, transc. \\
    \cite{donley2021easycom} \\ 
    \textbf{Tragic Talkers} & \Checkmark & 16($\times$2)$^{c}$ & 48 & 11($\times$2) & 2448$\times$2048 & \Checkmark & \Checkmark & \Checkmark & VA, 2D face bbox, 3D mouth pos.$^{a}$, \\
    &&&&&&&&& 2D pose keypoints$^{a}$, transc. \\
    \bottomrule
  \end{tabular}
  
  \begin{flushleft}
      \footnotesize{$^{a}$ Only a subset.}

      \footnotesize{$^{b}$ SSLR is mostly an audio-only dataset. We refer to the subset named `Human' that includes video recordings of human speakers.}
  
      \footnotesize{$^{c}$ In addition to two 16-element microphone arrays, audio is recorded with two lavalier microphones and a 4-channel FOA mic too.}
  \end{flushleft}
  
\end{table*}

\section{Introduction}

In 3D immersive production, there is little room for spatial misalignment between the audio and the visual displays. To provide realistic immersive experiences, accurate positioning of the audio sources with respect to the visual events that trigger them is therefore essential \cite{Stenzel:2018:PTC,Berghi:2020:IEEEVR}. With synthetic audio-visual assets, this is typically achieved through a manual process performed by the producer. The objects are located at the desired positions in the virtual environment enabling six degree-of-freedom (6-DoF) interaction of the end-user.
However, to achieve the same result and faithfully reproduce real-world scenes, an object-based audio-visual representation is required \cite{Coleman:2018:objectBased,pike2016object-based}. 
That is, object events must be extracted from the audio and visual streams and partnered with metadata describing specific attributes of the objects themselves, such as their content or positions in space.
An approach to enable real-world object-based media (OBM) production for 6-DoF interaction is by extracting the visual foreground objects and three-dimensionally reconstructing them. This is possible through a multi-view data capture \cite{Hartley:2004:multipleView}. Meanwhile, the acoustic sound-field of the audio events must be captured and tracked over time. This is possible through multi-channel audio, for example, recorded with a microphone array \cite{brandstein2001microphone}.

Producing such audio-visual content can be extremely costly as it requires expensive equipment and studio time. 
To the best of our knowledge, there are no available datasets with these characteristics. Existing datasets with multi-view video content usually neglect the audio component \cite{pumarola20193dpeople,CAPE:CVPR:20,Trumble:BMVC:2017}. On the other hand, datasets with multi-channel audio lack video content \cite{politis:2021:DCASE} or, when provided, the quality is rather low \cite{Lathoud:2004:AV16,He:2018:deepNetsHRI}.
Therefore, we propose Tragic Talkers: a sound- and light-fields dataset captured with the aid of two co-located audio-visual sensing devices. Co-locating audio and visual sensors is particularly useful to maintain the same fixed coordinate frame across the two modalities. 
The dataset consists of drama scenes, therefore, it is suitable for OBM production applications or OBM research. 
The actors were also provided with dedicated close microphones to record high-quality voice signals useful for reference and association when beamforming algorithms are applied.
Furthermore, we offer a range of curated labels to enable research in multiple domains, e.g., active speaker detection, speaker diarization, or body pose detection.
A preliminary version of the Tragic Talkers dataset was used in our work \cite{Berghi:2021:mmsp}. However, the previous version was smaller in size (20 sequences instead of 30) and was only labeled for 2D bounding boxes and voice activity.  

The rest of this paper is organized as follows: In chapter \ref{background_chap} existing datasets for 3D multimedia content are presented and their limitations discussed; In chapter \ref{setup_chap} the recording set-up employed to capture Tragic Talkers is described; In chapter \ref{design_chap} we motivate the design of the dataset and present the provided labels; In chapter \ref{uses_chap} few key uses for the dataset are presented; Finally, in chapter \ref{concl_chap} we conclude the paper.
Tragic Talkers is available for download at \url{https://cvssp.org/data/TragicTalkers/}. {The website provides ready-to-use PyTorch python scripts for data loading and audio-visual feature extraction.}

\begin{figure*}[bt]
  \centering
  \includegraphics[width=\textwidth]{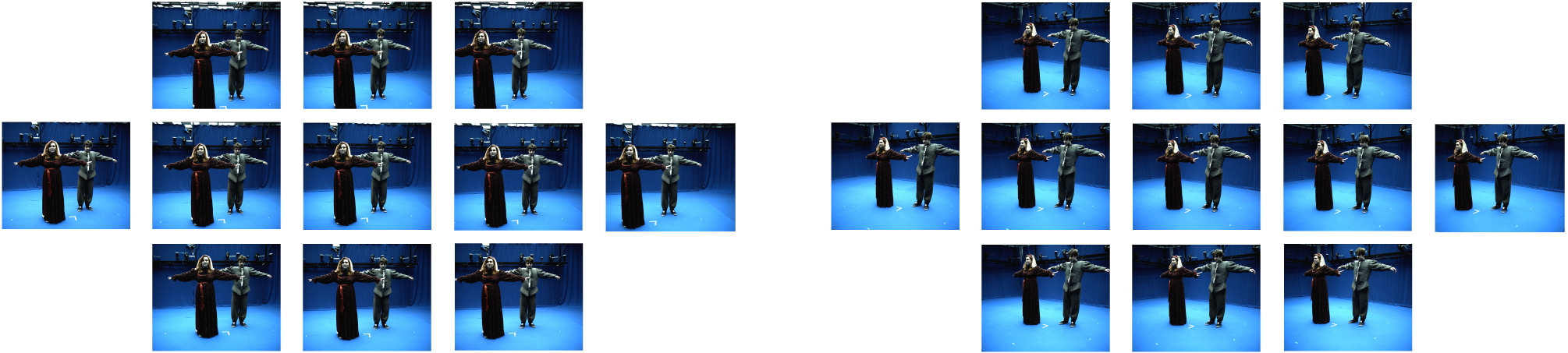}
  \caption{Same frame with actors in T-pose captured from the 11 cameras of AVA Rig 1 (right), and the 11 cameras of AVA Rig 2 (left).}
  \Description{The 22 points of view.}
  \label{view_points}
\end{figure*}

\section{Background} \label{background_chap}

Datasets for 3D audio-visual content production and research are important for a number of reasons. First of all, they represent a useful benchmark for evaluating new algorithms. Secondly, producing new multi-view and multi-channel audio-visual content can be extremely costly and time-consuming. Additionally, consistently-captured data can be employed in machine learning algorithms training.
Existing datasets that offer 3D multimedia content are typically designed to serve a specific technical aspect. 
3DPeople \cite{pumarola20193dpeople}, CAPE \cite{CAPE:CVPR:20}, SIZER \cite{tiwari20sizer}, and Deep Fashion3D \cite{zhu2020deep} are examples of real and synthetic datasets for 3D clothing modeling.
AVAMVG \cite{AVAMVG} is a multi-view dataset for 3D gait recognition.
TotalCapture \cite{Trumble:BMVC:2017} provides multi-view video data for 3D pose estimation.
However, none of the aforementioned datasets provide audio data. 
Volumetric datasets that also recorded the audio component are HUMAN4D \cite{Chatzitofis:2020:Human4D}, MHAD \cite{Ofli:MHAD}, CMUPanoptic \cite{Joo:CMUPanoptic}, and NAVVS \cite{Stenzel:IEEEVR:2021}, even though only MHAD and NAVVS released the audio recordings.
Even in these rare exceptions, the audio sources are recorded with dedicated microphones and not with the intention of capturing the directional cues of the acoustic sound-field of the performance, as it would be possible with microphone arrays or binaural microphones.
To the best of our knowledge, Tragic Talkers is the first dataset that simultaneously provides sound- and light-field video content.

In the literature, there exists a variety of datasets captured with microphone arrays for sound source localization and tracking. Some of them are summarized in \cite{roman2021micarraylib}.
In these cases, however, they lack visual content.
A good amount of multi-channel audio datasets that also include the visual modality are produced by the Inria group\footnote{https://www.inria.fr/en} for audio-visual tracking research, including the AVASM dataset \cite{Deleforge:2015:co-loc}, AVDIAR \cite{Gebru:2018:SpeakerDiarization}, RAVEL \cite{Alameda-Ravel2013}, and CAVA \cite{Arnaud:2008:CAVA}. Qian \textit{et al.} \cite{Qian:2019:AVSensDevic} propose a summary of these and other datasets.
However, in most cases, the resolution of the video stream is very low. Additionally, they are primarily recorded with the intent of testing speaker tracking algorithms. Therefore, the actors are asked to perform tasks such as counting, talking on top of each other, and in some cases improvising a conversation. 
In contrast, Tragic Talkers offers high-quality video content captured in a studio with the help of two student actors. The actors play famous lines from the tragedy of Romeo and Juliet and their costumes are chosen to resemble the clothing of the Italian Renaissance period.  
All these features representing the quality of performance and integrating stagecraft skill, method, and material, are typically referred to by the film industry with the term \textit{production value}.

Similar to \cite{Qian:2019:AVSensDevic}, in Table \ref{tab:datasets} we propose a summary of audio-visual datasets captured with microphone arrays for speech-related research. We omit some of the obsolete ones and include a few that have been released recently. 
In the table, a column has been reserved for the employment of actors. Excluding Tragic Talkers, EasyCom \cite{donley2021easycom} is the only dataset that proposes an acting performance as it presents groups of people talking at a shammed restaurant table ordering food from an actor restaurant server that collects the preferences and describes the dishes.  
AV16.3 \cite{Lathoud:2004:AV16} is the only dataset of the list that was not captured from a co-located platform but with three cameras located in different positions of the room and their resolution is fairly low. In the other datasets, the visual component is either captured with a single camera \cite{He:2018:deepNetsHRI,Qian:2019:AVSensDevic,donley2021easycom} or with a single stereo pair \cite{Arnaud:2008:CAVA,Gebru:2018:SpeakerDiarization}.
Tragic Talkers is the only dataset that is captured with two separate sets of 11 high resolution cameras for light-field video production. Additionally, Tragic Talkers is captured against a chroma-consistent background facilitating silhouette extraction.
Excluding EasyCom \cite{donley2021easycom}, all the datasets are quite small and therefore unsuitable for learning-based research. Tragic Talkers can be either seen as a 30-sequences dataset captured from 22 different points of view or, considering the two rigs separately, as a 60-sequences dataset captured with a single rig. 
Viewed this way, it offers 3.8 hours of total annotated video data with audio that is semantically, temporally and spatially coherent.
Therefore, we believe it is of sufficient scale as to enable a range of machine-learning research.




\section{Capturing Set-up} \label{setup_chap}

Tragic Talkers was captured with the aid of two twin Audio-Visual Array (AVA) Rigs. Each AVA Rig is a custom device consisting of a 16-element microphone array and 11 cameras fixed on a flat perspex sheet supported by a tripod. Therefore, the AVA Rig represents a co-located multi-sensing platform. That is, each sensor is in a fixed relative position with respect to the other sensors of the array and the same coordinate frame is shared across the modalities.
The array has size 0.6m$\times$0.4m and is compact and lightweight enough to be mounted on a standard production tripod. 
The 11 cameras (Grasshopper3 USB3\footnote{https://www.flir.com/products/grasshopper3-usb3/?model=GS3-U3-51S5C-C}) capture synchronized 2448$\times$2048p video frames at 30\,fps.
They are positioned in three rows and are equally spaced from one another, with the exception of the central camera. 
The camera array has been designed to support 6-DoF content production for the limited volume of a seated user.
Before the capture session, intrinsics and extrinsics of the camera arrays were calibrated using a chart \cite{zhang:1999:calib,Hartley:2004:multipleView}. Therefore, the two camera rigs share the same world reference coordinated with the origin being the central camera of AVA Rig 1.   
Audio is sampled at 48 kHz, 24 bits. The microphone array can be used for sound source localization. It has a horizontal aperture of 450\,mm and a vertical aperture of only 40\,mm. This results in higher resolution when localizing audio sources along the azimuth direction than elevation, which is consistent with human perception \cite{Strybel:2000:MinimumAA}.
The horizontal design of the microphone array is log-spaced to focus on broad frequency coverage from 500\,Hz to 8\,kHz, to better support the horizontal speech band resolution.
After the capture session, audio recordings have been channel-wisely multiplied by gain factors to set their root mean square (RMS) magnitude to a common value and achieve level calibration, as in \cite{vasudevan:2020:semanticobject}.  

During the data capture, the two AVA Rigs were placed side by side at a distance of about 2m from one another to provide alternative view points and emulate a plausible camera arrangement in a standard video production set, generating an angle of about 30° to 50° between the rigs and each subject.  
In addition to the acoustic sound-field captured by the microphone arrays, actors were accoutered with two lavalier microphones to record high-quality voice signals and a first-order ambisonic room microphone (Soundfield DSF1\footnote{https://www.soundfield.com/\#/products/dsf1}) useful for reverb estimation was placed above the capturing volume. 
A clapper-board was employed at the beginning and at the end of each take to enable manual alignment of the audio and visual streams.
The actors were asked to start and conclude their performance with the T-pose. 
The entire dataset was captured against a blue background to facilitate silhouette extraction by chroma-keying and 3D reconstruction using camera calibration data. Background images of the empty studio captured with the AVA Rigs are provided.


\section{Design of Dataset} \label{design_chap}

The Tragic Talkers dataset is designed to suit a variety of research areas including, but not limited to, light-field and sound-field processing. The intention to generate high quality content for next-generation 3D audio-visual production led to the decision of employing semi-professional actors with predefined scripts and selected costumes. Therefore, we chose to capture excerpts from Shakespeare's Romeo and Juliet drama, hence the name `Tragic Talkers', where student actors played the role of the two key characters. In the rest of this paper, we will refer to `Juliet' as the actress who interpreted Juliet and, similarly, to `Romeo' as the actor who interpreted Romeo.
Tragic Talkers consists of 30 sequences ranging from about 13 to 43 seconds in length (21 seconds on average), yielding over 400K frames (3.8 hours) of video data considering the 22 separate views.
In order to produce a diverse set of talking conditions useful to speech-related research, three different types of scenes were recorded: \textit{monologues} with single-character oration performance to provide male-only and female-only speech content; \textit{conversations} with the two actors standing side-by-side exchanging lines from the famous tragedy; and \textit{interactive} scenes where the actors were asked to move within the capture volume while talking, occluding each other and improvising their conversation.  
The last provides more challenging content for applications such as 2D and 3D audio-visual tracking.
Although the dataset does not include scenarios with multiple simultaneous voices, the nature of the recordings allows synthetic creation of audio-visual cocktail parties by combining speech sequences: the audio is mixed, while the respective actors' video extracted from the background using silhouettes is overlaid to create augmented multi-talker sequences.

\begin{table}[bt]
  \caption{Speech activity over the dataset's 30 sequences.}
  \label{tab:speechAct}
  \begin{tabular}{c|c|rr|rr}
    \toprule
    Partition & Duration & Active & Silent & Romeo & Juliet \\
    \midrule
    Development & 519s & 271s & 248s & 116s & 155s \\
    set && 52\% & 48\% & 43\% & 57\% \\ \hline
    Test & 121s & 64s & 57s & 31s & 33s \\
    set && 53\% & 47\% & 48\% & 52\% \\ \hline
    Overall & 640s & 335s & 305s & 147s & 188s \\
    && 52\% & 48\% & 44\% & 56\% \\
    
    \bottomrule
  \end{tabular}
\end{table}

\begin{figure}[bt]
  \centering
  \includegraphics[width=\linewidth]{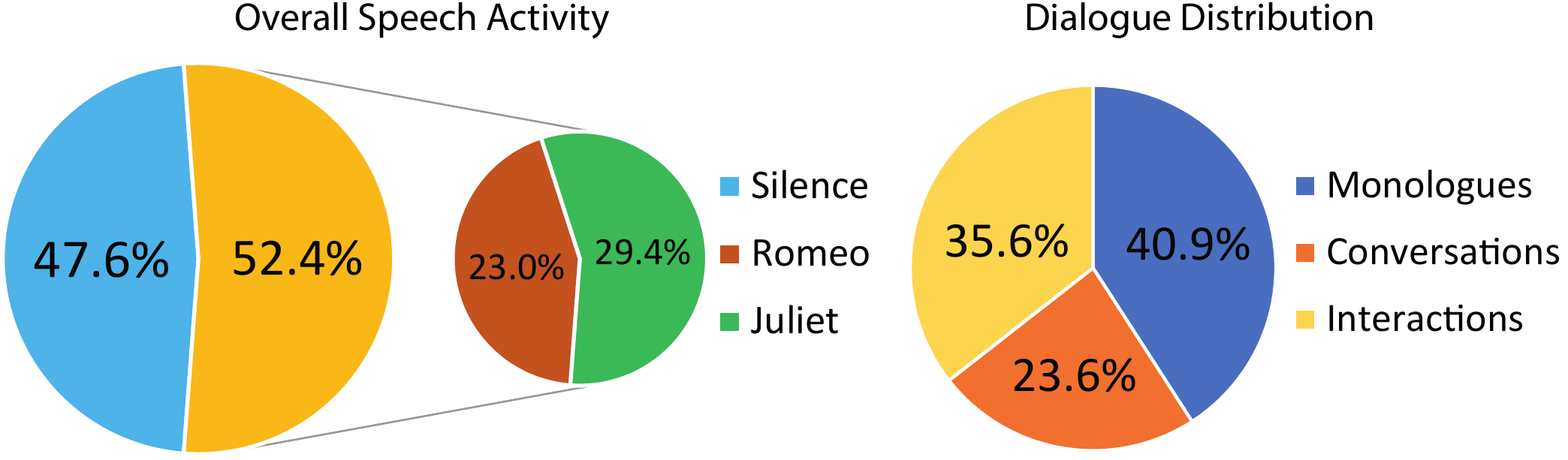}
  \caption{On the left, a pie chart showing the overall distribution of speech and silence across the dataset and how the speech content is distributed between Romeo and Juliet. On the right, the distribution of monologues, conversations, and interactions.}
  \Description{Pie charts showing the speech activity and the dialogues distributions.}
  \label{pie_charts}
\end{figure}

\begin{figure}[bt]
  \centering
  \includegraphics[width=\linewidth]{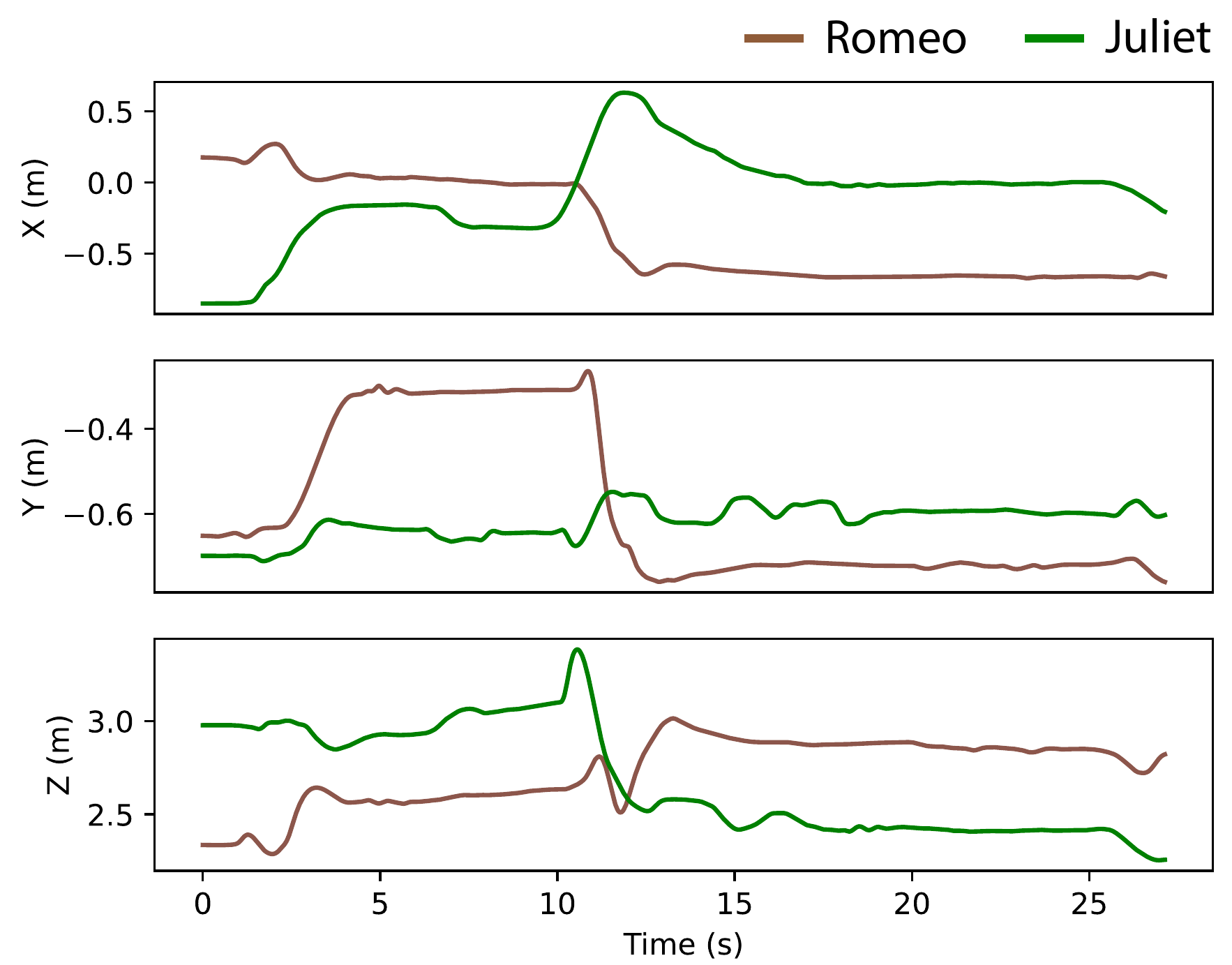}
  \caption{3D mouth tracking results for Romeo and Juliet in the sequence `interactive1$\_$t2' of the test set.}
  \Description{X, Y, Z mouth detections over time for Romeo and Juliet.}
  \label{3DTrack}
\end{figure}

\begin{figure*}[bt]
  \centering
  \includegraphics[width=\textwidth]{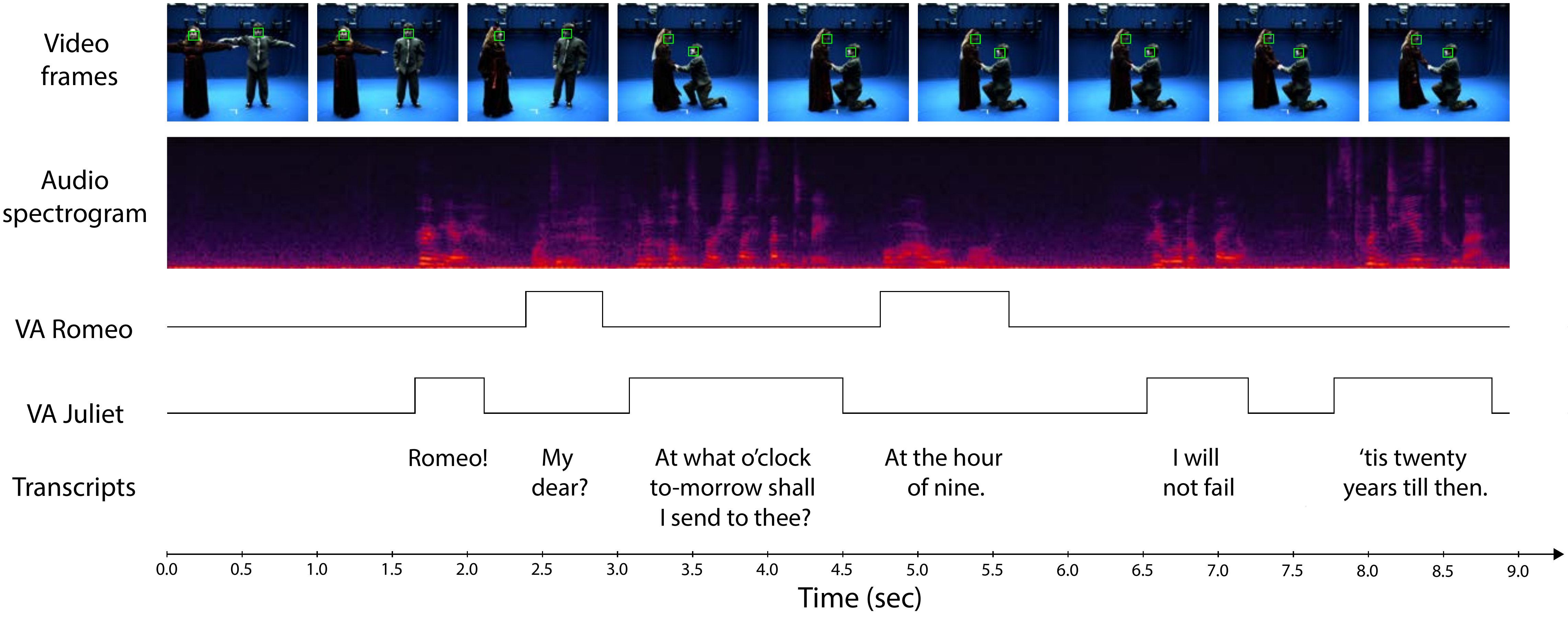}
  {\mbox{ }\hspace{4.5ex}\includegraphics[width=162mm]{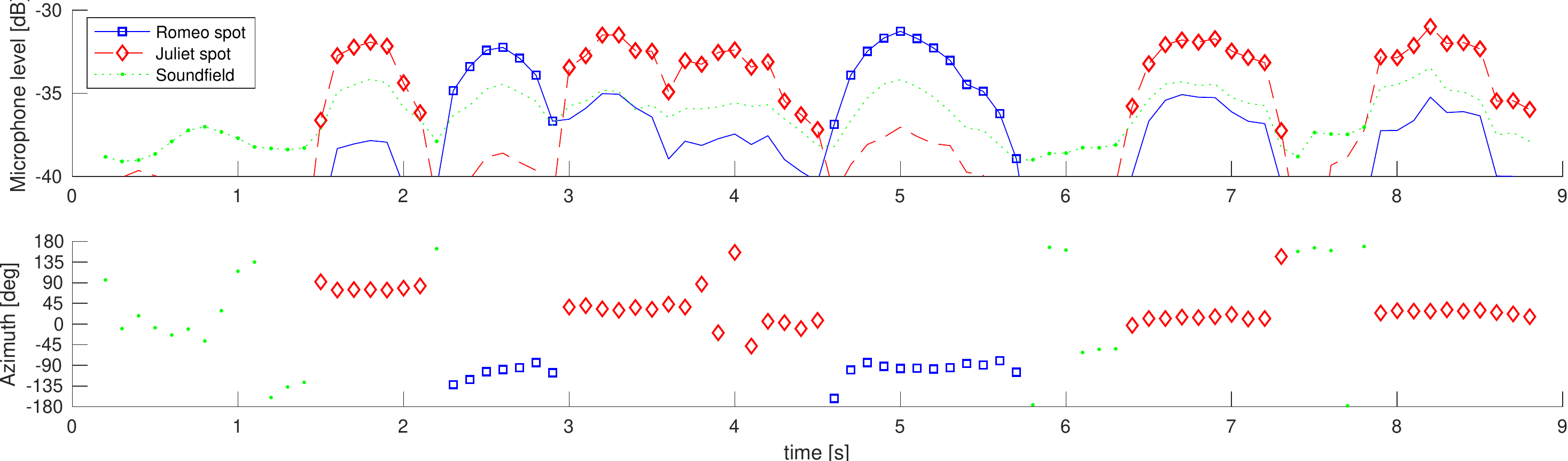}}
  \caption{Audio-visual content of an excerpt from the sequence `interactive1$\_$t2' depicted by video frames overlaid with detected-face bounding boxes (reported at 1 fps) and audio spectrogram. Alongside are voice activity (VA) labels for Romeo and Juliet, and the speech transcriptions over time.
  The middle axes below show normalized sound levels at spot microphones worn by Romeo and Juliet, which correspond with the voice activity labels above, and the soundfield microphone mounted in the room.
  The bottom axes illustrate the active speakers' azimuth angle in relation to the soundfield microphone, exploiting the directivity of the first-order-ambisonic audio signals.}
  \Description{Figure illustrating video frames, face bounding boxes, audio spectrogram, voice activity labels, and transcriptions over time.}
  \label{VA_labels}
\end{figure*}

\subsection{Labels} \label{labels_chap}

Along with the audio-visual data, we provide an extensive set of labels useful to test a variety of algorithms. Here, we summarize the available labels and explain how they have been generated.
First of all, we have manually produced the voice activity (VA) labels with actor ID association. In order to achieve that, we employed the software Adobe Audition\footnote{https://www.adobe.com/products/audition.html} to manually find the onset and offset of each speech segment as, alongside the audio tracks, it also provides the visual reference of the spectrograms. 
Table \ref{tab:speechAct} reports the distribution between speech and silence across the dataset, as well as the distribution of Romeo's speech and Juliet's speech.
Secondly, we provide the sequence transcriptions with the respective speaker ID. To generate them, we employed Otter.ai\footnote{https://otter.ai/}
Then, we manually adjusted the incorrect transcriptions.

To enable speaker localization and tracking, 2D face bounding boxes for each camera view are provided. Face detection was performed on each video frame with S3FD \cite{Zhang2017S3FDS}. When the face detector failed, for instance when the actors are not facing the camera, a bounding box was drawn by a manual educated guess. In this way, the actors are continuously tracked in the 2D domain throughout the entire duration of the sequence.
Face tracks were generated from the per-frame detections based on intersection over union (IoU) across adjacent frames. Then, the corners of the bounding boxes have been interpolated over time to provide smooth temporal coherence.
In addition to that, we selected 5 sequences from the dataset to constitute a test set. In order to be heterogeneous and representative of the dataset, the test set includes two monologues, one from Romeo and one from Juliet, one conversation, and two interactive sequences. The test set represents a useful benchmark for learning-based algorithms and it was labeled for 3D mouth position.
To generate the 3D mouth labels, we employed OpenPose \cite{Cao:2029:openpose} to
detect 2D skeletal keypoints of the two actors on the 2D video frames.
In order to provide spatial and temporal coherence to accurately track the 3D location of the mouth joint, the 2D detections have been sorted as described in \cite{Malleson:2020:realtime}. 
The sorted 2D detections are combined with confidences to
remove the impact of potentially unreliable detections and estimate the 3D locations via camera triangulation.
2D skeleton keypoints detected with OpenPose are provided. Examples of 3D mouth detections over time from one of the interactive sequences are reported in Figure \ref{3DTrack}.

\begin{figure}[bt]
  \centering
  \includegraphics[width=\linewidth]{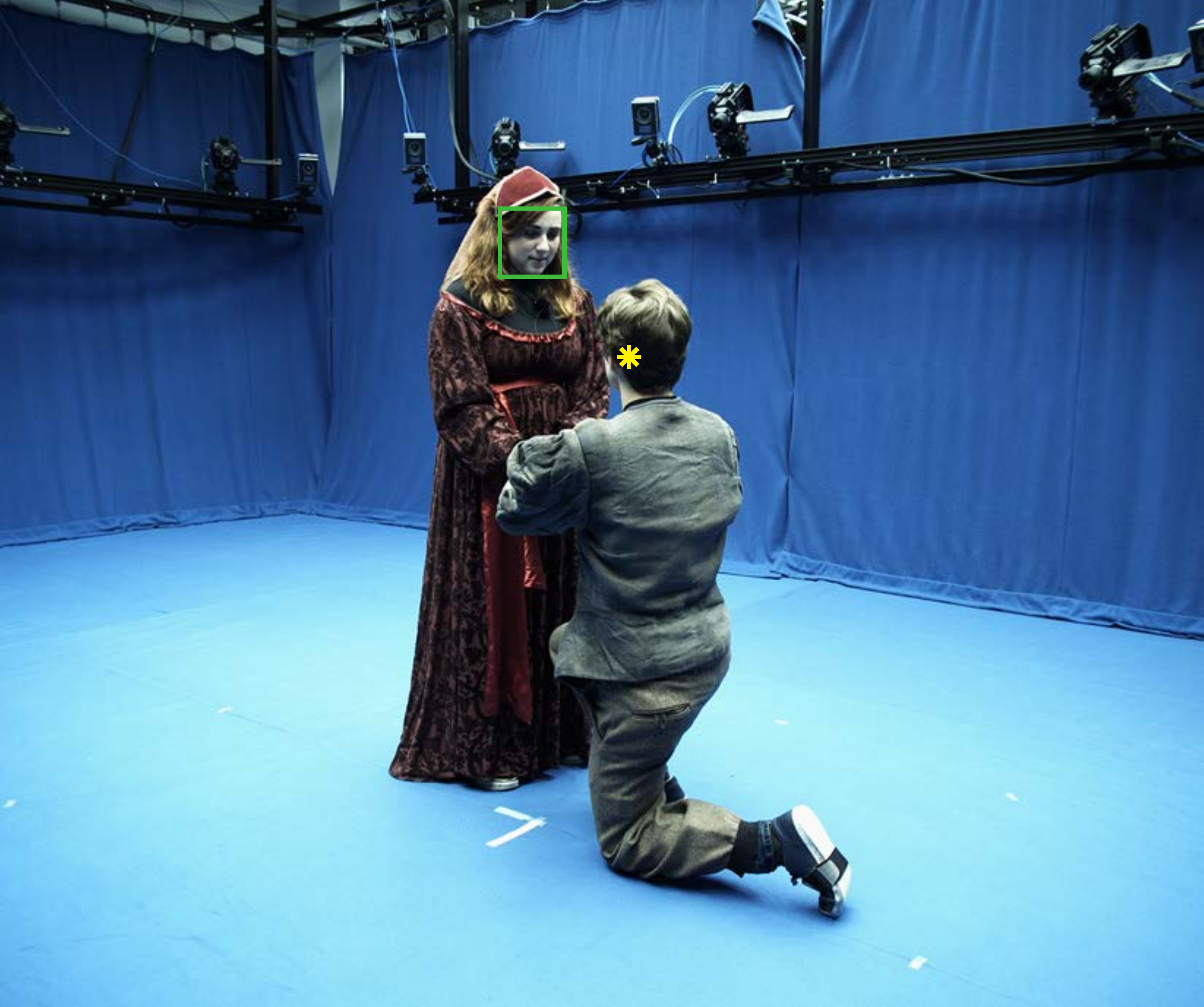}
  \includegraphics[width=\linewidth]{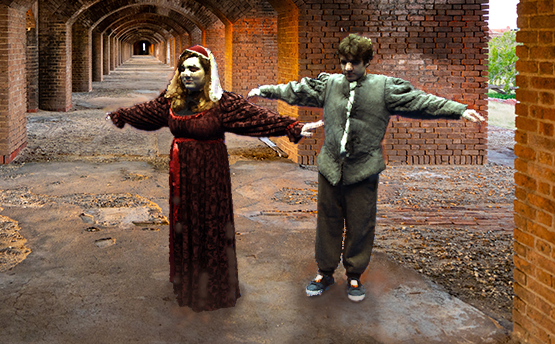}
  \caption{Examples of possible uses of the dataset. In the top picture, it is proposed an example of audio-visual active speaker tracking application, where the face of Juliet is detected by a face detector while the voice of Romeo is correctly located by the microphone array (yellow asterisk), even though his face is not visible.   
  Below, an example of a partial 3D reconstruction of Romeo and Juliet positioned in a background of our choice.}
  \Description{Two images are shown. One represents Romeo on his knees in front of Juliet in our studio. It depicts a possible audio-visual speaker tracking scenario. The other is a photo composition of Romeo and Juliet partially reconstructed from the light-field video stream and located in a different background. The chosen background portrays arcades made of bricks. Romeo and Juliet are standing in T-pose.}
  \label{uses_examples}
\end{figure}

\section{Uses of the dataset} \label{uses_chap}

Tragic Talkers can be employed for multiple applications. Here we propose some of the key use cases for this dataset.
The availability of 2D and 3D tracking data clearly suggests the employment of this dataset in speaker localization and tracking applications. This can either be addressed with audio-only algorithms, for instance by applying beamforming \cite{galindo:2020:microphone} or SRP-PHAT  \cite{Do:2007:RT_SRP-PHAT}, or with audio-visual solutions, as in \cite{Qian:2019:AVSensDevic}.
Tragic Talkers enables object-based spatial audio production through the generation of tracking metadata and automatic association with the high-quality signals from the additional microphones provided. For instance, \cite{izhar:2020:AVtracker} proposed to generate automatic 3D spatial metadata with an audio-visual tracker and utilize it to drive a spatial beamformer to steer the listening directivity of the microphone array towards the predicted positions of the audio sources. The beamformed audio signals are then automatically associated to, and replaced by, the high-quality recording spatialized at the predicted positions. 

The large amount of views with which the dataset was captured enables learning-based audio-visual research for 2D active speaker detection with multi-channel audio signals \cite{Berghi:2021:mmsp,donley2021easycom}. In fact, over 3.8 hours of monocular video content can be generated from the 30 sequences provided, all accompanied by multi-channel audio recordings.
For instance, in our earlier work \cite{Berghi:2021:mmsp} where a preliminary version of the Tragic Talkers dataset was employed, we adopted a student-teacher learning pipeline. We proposed to generate pseudo-labels from the video sequences with a pre-trained active speaker detector (teacher) and used them to train a multi-channel audio network (student) to detect and regress the position of the speakers directly in the video frames.
The availability of the dialogue transcriptions also allows the speaker diarization task to be performed \cite{Gebru:2018:SpeakerDiarization}.
The light-field camera array on each of the two AVA Rigs allows 3D dynamic foreground character reconstruction for immersive production.
For example, the central camera of the AVA Rigs can be employed as a reference view for an operator during the editing and compositing of the experience. The surrounding cameras can be used for automatic partial 3D reconstruction of Romeo or Juliet. 
The size of the aperture of the camera array enables 6-DoF movement in the limited volume of a seated user. The partially reconstructed foreground avatars can then be inserted into a virtual environment. An example of 3D reconstruction of Romeo and Juliet inserted in a background location of our choice is proposed in the lower image of Figure \ref{uses_examples}. Furthermore, the 3D avatars can be accompanied by their respective voices, spatialized as previously described, to produce an object-based audio-visual immersive experience, as suggested in \cite{Schweiger:2022:tool6dof}.

The bi-modal multi-view content offered in Tragic Talkers is optimal to evaluate rendering algorithms too. 
In the context of audio-visual text-to-video synthesis (see, for instance, Synthesia\footnote{https://www.synthesia.io/}), Tragic Talkers can be employed as an evaluation set to test audio-visual speech rendering algorithms from different viewpoints.
This could study different view point interpolation or replace one character with an avatar, and supports formal subjective experiments for perceptual quality evaluation.
Another potential use of the data involves the construction of synthetic datasets, for instance, by replacing the background with synthetic locations \cite{Caliskan:2020}, or combining the sequences to create different combinations of multi-talker scenarios, which can be useful as part of a data augmentation strategy. 
In addition to these tasks, we envisage a contribution to fundamental audio-visual machine learning may be explored through the use of these data.

\section{Conclusion} \label{concl_chap}

This paper introduces the Tragic Talkers dataset, the first sound-field and light-field dataset, produced with drama students performing excerpts from Shakespeare's famous tragedy, Romeo and Juliet.
Tragic Talkers was captured with the aid of two twin AVA Rigs, each consisting of an 11-element light-field camera array and a 16-element microphone array. High-quality speech data recorded with two lavalier spot microphones as well as a first-order ambisonics room microphone are also provided.
The dataset is labeled for voice activity, 2D face bounding boxes, 3D mouth coordinates, and dialogue transcriptions. We discussed potential uses of the dataset, including 2D or 3D speaker tracking, object-based media production, and light-field 3D reconstruction for 6-DoF interaction. 

In future work, we aim to use Tragic Talkers to investigate the effect of different temporal, spatial and semantic aspects of audio-visual coherence \cite{Arandjelovic:2017:look,Afouras_2022_CVPR}.

\begin{acks}
This work is supported by InnovateUK (105168) `Polymersive: Immersive video production tools for studio and live events' and a PhD studentship from the Doctoral College of the University of Surrey. For the purpose of open access, the authors have applied a Creative Commons Attribution (CC BY) license to any Author Accepted Manuscript version arising. 
Data supporting this study are available from our web server (DOI 10.15126/surreydata.900446, \url{https://cvssp.org/data/TragicTalkers}). 
Thanks to actors Phoebe Salem and Mason Stickland for their time and availability; 
to Mohd Azri Mohd Izhar, Hansung Kim, and Charles Malleson for their contribution to the audio-visual recordings;
to Umar Marikkar for developing utility scripts supporting easy access to the data, e.g., data loaders and audio-visual feature extractors;
to Alexander Todd for helping during the laser cut of the AVA Rigs.
\end{acks}

\bibliographystyle{ACM-Reference-Format}
\bibliography{CVMP22}


\end{document}